# Substrate-engineered tunable bound states in the continuum and directional radiation in dielectric metasurfaces

Hao Song[1,2]*, Yanming Sun[3], Jian Li[1,2], Wanlin Wang[4], Ming Chun Tang[5]

[1]*College of Physics and Electronic Information Engineering, Neijiang Normal University, Neijiang, 641100, China*

[2] *Neijiang Optoelectronic Devices Engineering Research Center, Neijiang, 641100, China*

[3] *State Key Laboratory of Radio Frequency Heterogeneous Integration, College of Physics and Optoelectronic Engineering, Shenzhen University, Shenzhen 518060, China*

[4]*School of Physical Sciences, Great Bay University, Dongguan, 523000, China*

[5]*College of Microelectronics and Communication Engineering, Chongqing University, Chongqing 400044, China*

*Email: *haosongnju@njtc.edu.cn*

**Abstract**

Tunable bound states in the continuum (BICs) in metasurfaces offer powerful opportunities to control light-matter interactions, yet the role of out-of-plane symmetry breaking remains poorly understood. Here, we reveal a mechanism that enables tunable high-$Q$ BICs and directional radiation through out-of-plane symmetry breaking in all-dielectric metasurfaces. A substrate-free metasurface composed of periodically arranged multilayer cylinders that support overlapping magnetic dipole and electric quadrupole resonances, yielding electric mirror and symmetry-protected BIC responses at 1550 nm. Introducing multilayer substrates breaks out-of-plane symmetry and excites guided modes. When the guided-mode wavelength matches that of the BIC and coupling to the substrate is suppressed, the BIC wavelength remains nearly invariant, while the $Q$ factor increases with layer number. In contrast, spectral detuning and enhanced coupling lead to pronounced blueshifts and rapid $Q$ degradation. The interplay between guided-mode matching and coupling strength thus governs whether a BIC remains robust or becomes tunable. These findings establish a general framework for BIC engineering via out-of-plane symmetry breaking and provide a versatile platform for tunable metasurfaces with potential applications in integrated optics.

## 1 introduction

Optical bound states in the continuum (BICs) are non-radiative, spatially localized states embedded within the continuum of radiation modes above the light line in momentum space [1]. These states confine light in micro- and nanostructures with, in principle, infinitely high quality factors ($Q$), as all radiative loss channels are suppressed [2, 3]. This property is crucial for enhancing light-matter interactions and enabling advanced photonic functionalities. In recent years, extensive efforts have been devoted to BIC-supported all-dielectric metasurfaces, with demonstrated applications in lasing [4], absorption enhancement [5], chiral sensing [6], and directional emission [7], among others. Metasurfaces are two-dimensional (2D) metamaterials composed of periodic meta-atoms, offering a bottom-up design paradigm in which tailored functional responses are achieved through precise control of individual elements [8]. Therefore, all-dielectric metasurfaces provide several key advantages for realizing and manipulating BICs. First, they offer substantial design flexibility in controlling the amplitude, phase, polarization, and radiation properties of electromagnetic waves through geometric tuning, material selection, and spatial arrangement of meta-atoms [8, 9]. Second, the absence of intrinsic non-radiative absorption losses in dielectric materials makes these platforms ideal for achieving BICs by selectively engineering radiative losses [2, 10]. Third, their low-profile and ultrathin characteristics make them highly suitable for integration into compact optoelectronic systems. To date, a wide range of studies on BICs in all-dielectric metasurfaces have demonstrated promising applications in chiral optics [6, 11], nonlinear optics [12, 13], directional emission [7, 14], sensing [15], light absorption [16], and temporal photonics [17]. More recently, the sensitivity of BIC quality factors to positional perturbations has been exploited to overcome diffraction limits and enable nanoscale alignment, offering new opportunities for high-precision metasurface fabrication [18, 19].

In momentum space ($k$-space), BICs are generally classified into symmetry-protected BICs at the Γ point and accidental BICs located away from the Γ point [20]. These states correspond to far-field polarization singularities characterized by integer topological charges [21]. Accidental BICs originate from the destructive interference of resonant modes, which suppresses all outward radiation channels and leads to mode confinement. Upon continuous structural perturbation, an off-Γ BIC evolves into a pair of circularly polarized states with opposite chirality, each carrying a half-integer topological charge [22, 23]. In contrast, symmetry-protected BICs can be readily realized in dielectric metasurfaces possessing in-plane symmetry. However, their non-radiative nature prevents direct coupling to the far field, thereby limiting their practical utility. To overcome this limitation, quasi-BICs can be engineered from ideal BICs through structural perturbations or oblique incidence, enabling controlled radiation leakage and accessible far-field responses [24, 25]. Generally, the $Q$ value of the quasi-BICs decays as a power function of the perturbations. To date, numerous studies have achieved quasi-BICs by introducing in-plane symmetry breaking in metasurfaces via asymmetric, rotated, or tilted meta-atom designs, leading to a wide range of applications in linear and nonlinear photonics [5-7, 11, 12, 15-17, 23, 26-30].

An important question concerns the role of out-of-plane symmetry breaking in BICs supported by dielectric metasurfaces. Previous studies suggest that achieving high-$Q$ BICs generally requires the preservation of out-of-plane symmetry [31]. One approach to breaking this symmetry involves introducing structural asymmetry along the $z$-direction, for example, by designing meta-atoms with unequal thicknesses. By tuning the thickness contrast, such metasurfaces can induce a transition from ideal BICs to quasi-BICs [32]. In addition, several studies have explored the combined effects of out-of-plane and in-plane symmetry breaking in meta-atoms to achieve similar transitions [23, 33, 34]. Another common strategy is to introduce a substrate to achieve the out-of-plane symmetry-breaking metasurfaces. In most practical cases, metasurfaces are supported by homogeneous dielectric substrates, which primarily serve as mechanical supports and optical spacers [30, 32, 34, 35]. Accordingly, symmetry breaking at the level of individual meta-atoms is often considered the dominant mechanism governing the transition from BIC to quasi-BIC. However, the specific role of the substrate in influencing BIC responses remains insufficiently understood. More recently, efforts have been devoted to substrate engineering, particularly through the use of multilayer structures composed of different materials. These platforms enable simultaneous manipulation of BICs and guided modes, giving rise to phenomena such as enhanced Goos-Hänchen shifts [36], polariton BIC solitons mediated by polariton-polariton interactions [37], and strong coupling effects between high-reflectivity substrates and BICs for applications in chiral detection [6] and perfect absorption [26]. However, despite these advances, it remains an open question how multilayer substrates influence BICs in dielectric metasurfaces and whether high-$Q$ BICs can be preserved in the presence of out-of-plane symmetry breaking.

Here, we investigate the mechanism of substrate-enabled tunable BICs in an all-dielectric metasurface, where a carefully designed multilayer substrate is employed to break out-of-plane symmetry. Based on Mie resonances and anisotropic two-dimensional materials, we adopt a bottom-up design strategy to realize a substrate-free metasurface composed of periodically arranged meta-atoms supporting overlapping scattering resonances, thereby enabling both BIC and electric mirror responses at 1550 nm [38-40]. We then introduce three types of multilayer substrates constructed from alternating high-index-contrast materials. By tuning the thickness of individual layers, controlled excitation of guided modes can be achieved. Notably, when the guided-mode wavelength matches the operating wavelength of the BIC and coupling between the substrate and meta-atoms is suppressed, the substrate has a negligible impact on the BIC response, resulting in an enhanced $Q$. In contrast, when wavelength detuning and coupling are present, the system exhibits tunable BIC responses with a pronounced dependence on the number of layers, accompanied by a reduction in $Q$. Furthermore, the metasurface demonstrates layer-number-dependent unidirectional radiation away from the Γ point in momentum space, arising from the interplay between guided modes and coupling effects. These findings introduce an additional degree of freedom for tailoring tunable BICs and directional radiation in metasurfaces, provide new insights into BIC mechanisms under out-of-plane symmetry breaking, and pave the way for practical applications in integrated optics and photonics.

## 2 Results and Discussions

### 2.1 Scattering resonance overlapping

The overlap of Mie resonances gives rise to unique optical phenomena, such as Kerker scattering, which enables unidirectional forward or backward scattering [38, 41]. In this work, we design a tailored meta-atom comprising an infinitely long multilayer cylinder of $MoSe_2$ and $SiO_2$, as schematically illustrated in Figure 1(a). The refractive index ($n_l$) of $SiO_2$ is taken as 1.45 [42]. $MoSe_2$, a transition metal dichalcogenide (TMD), exhibits pronounced optical anisotropy. At an incident wavelength of 1550 nm, the radial ($n_r$) and azimuthal ($n_t$) refractive indices are approximately 4.22 and 2.78, respectively [43]. For simplicity, all $MoSe_2$ layers are assumed to have identical thicknesses, and the same applies to the $SiO_2$ layers. These layers are alternately arranged along the radial direction, with individual thicknesses much smaller than the operating wavelength, thereby satisfying the long-wavelength approximation. Under this condition, the effective radial and azimuthal refractive indices can be described using effective medium theory (EMT) as follows [44]

$$n_{te} = \sqrt{f_1 n_l^2 + (1 - f_1) n_t^2} \ , \tag{1}$$

and

$$n_{re} = \frac{n_l n_r}{\sqrt{(1 - f_1) n_l^2 + f_1 n_r^2}} \ . \tag{2}$$

Here, $f_l$ is the volume ratio of $SiO_2$ to the total cylinder. The anisotropic parameter is defined as $\eta = n_{te}/n_{re}$ .

Next, we analyze the scattering properties of the cylinder embedded in air. A transverse electric (TE) wave with wavevector **k** propagates perpendicular to the cylinder axis, with its magnetic field **H** oriented along the axis. The analytical treatment of Mie scattering begins with Maxwell's equations and the application of appropriate boundary conditions. The resulting *m*th-order scattering coefficients can then be expressed as follows [44, 45]

$$a_m = \frac{n_{te} J_{\tilde{m}}(n_{te} x_r) J'_m(x_r) - J_m(x_r) J'_{\tilde{m}}(n_{te} x_r)}{n_{te} J_{\tilde{m}}(n_{te} x_r) H_m^{(1)\prime}(x_r) - J'_{\tilde{m}}(n_{te} x_r) H_m^{(1)}(x_r)} \ , \tag{3}$$

where the size parameter $x_r$ equals wavenumber $k$ times radius $r$, the radial anisotropy-revised function order $\tilde{m} = m\eta$, $J_m(x)$ and $H_m(x)$ are the first kind of Bessel and Hankel functions, respectively. After normalization by the geometric cross-section and the single-channel limit $2\lambda/\pi$ [46], the scattering efficiency is written by [47]

$$N_{sca} = |a_0|^2 + 2\sum_{m=1} |a_m|^2 \ . \tag{4}$$

According to the cylindrical coordinate symmetry, the *0*th order denotes the magnetic dipole (MD), the *1*st order is the electric dipole (ED), and the *2*nd order represents the electric quadrupole (EQ) [48].

The scattering responses of the multilayer cylinder with $f_l$=0.4805 are presented in Figure 1(b). The red curve represents the total scattering efficiency, while the solid blue, green, and black curves correspond to the multipolar contributions of the magnetic dipole (MD), electric dipole (ED), and electric quadrupole (EQ), respectively. Notably, the MD and EQ modes exhibit coincident resonances at $x_r$ = 2.06, whereas the contribution of the ED mode is negligible. To further validate these results, we perform numerical simulations using the finite element method implemented in COMSOL Multiphysics [49]. In the model, the cylinder is approximated as a 2D cross-sectional multilayer circle suspended in air, with the surrounding domain terminated by perfectly matched layer (PML) boundaries. The blue circles denote the total normalized scattering efficiencies obtained from simulations, showing excellent agreement with the analytical results (red curve). According to the EMT, the effective anisotropic refractive indices are $n_{te}$=3.2034, and $n_{re}$=1.8388, yielding an anisotropy factor $\eta$ of about 1.742. Under this approximation, the multilayer cylinder can be treated as an equivalent homogeneous anisotropic cylinder. The dashed blue curve represents the corresponding simulated scattering efficiency. Overall, the numerical and analytical results are in good agreement, confirming the resonance overlap at $x_r$=2.06.

Based on the scattering multipoles, the angular scattering spectra of a cylinder can be expressed as

$$I_{SA}(\theta) = \frac{2}{k\pi}\left|a_0 + 2\sum_{m=1}^{\infty} a_m \cos(m\theta)\right|^2 , \tag{5}$$

where $\theta$ is the scattering angle [44, 47]. To characterize the unidirectional scattering, a parameter $g$ is defined by [38, 47]

$$g = \frac{\oint_s \cos\theta I_{SA}(\theta) ds}{\oint_s I_{SA}(\theta) ds} . \tag{6}$$

Thus, unidirectional forward scattering ($\theta$=0◦) occurs when $g$=1, and unidirectional backward scattering (θ=180◦) occurs when $g$=-1. For our case, the angular scattering spectrum at $x_r$=2.06 is shown in Figure 1(c). The interference between the MD and EQ resonances leads to nearly equal intensities of forward and backward scattering because $g$=0.031, whereas the contribution from the ED is negligible. Our previous work has shown that enhanced backward scattering is beneficial for achieving high reflectivity in dielectric metasurfaces [39].

For comparison, we analyze the scattering properties of a $SiO_2$ cylinder with the same radius as the multilayer cylinder. According to Mie scattering theory [47], the $m$th-order scattering coefficient can be expressed as follows,

$$b_m = \frac{n_l J'_m(x_r) J_m(n_l x_r) - J_m(x_r) J'_m(n_l x_r)}{n_l J_m(n_l x_r) H_m^{(1)\prime}(x_r) - J'_m(n_l x_r) H_m^{(1)}(x_r)} . \tag{7}$$

By replacing $a_m$ with $b_m$ in equation (4), the scattering properties of the $SiO_2$ cylinder can be obtained. As shown in Figure 1(d), the red curve represents the total scattering efficiency, while the blue, green, and black curves correspond to the multipolar contributions of the MD, ED, and EQ, respectively. The results indicate that the $SiO_2$ cylinder does not support any pronounced resonant modes or resonance overlap. This conclusion is further confirmed by the numerical simulations, as indicated by the blue circles. Figure 1(e) shows the angular scattering spectrum of the $SiO_2$ cylinder at $x_r$=2.06. The forward scattering is dominant, with g=0.793, arising from the interference of non-resonant modes.

Figure 1(f) shows the scattering electric field magnitude $|\mathbf{E}_s|$ distribution of the effective cylinder at $x_r$=2.06 with the wavevector **k** propagating from left to right. The field profile is in good agreement with the analytically calculated distribution in Figure 1(c), exhibiting pronounced backward scattering. Then, Figure 1(g) presents the $|\mathbf{E}_s|$ distribution of the actual multilayer cylinder with 30 alternating layers. To clearly visualize the internal field distribution, the outer boundary is omitted. The resulting field profile agrees well with both the effective-medium model and the analytical calculations. For comparison, Figure 1(h) shows the $|\mathbf{E}_s|$ distribution of the $SiO_2$ cylinder, which exhibits dominant forward scattering, consistent with the analytical result in Figure 1(e). These results demonstrate that the $MoSe_2$-$SiO_2$ multilayer cylinder provides enhanced control over resonance overlap and enables strong backward scattering.

**2.2 Substrate-free metasurface**

Figure 2(a) depicts a schematic of a metasurface with out-of-plane symmetry, composed of periodically arranged multilayer cylinders operating at $x_r$=2.06. The cylinders are arranged with a minimum surface separation $dP$, yielding a lattice period $a=2r+dP$, where $r$ denotes the cylinder radius. Two wavevector configurations are considered. In the first configuration, an out-of-plane incident wave is normally incident on the metasurface. In the simulations, periodic boundary (PB) conditions are applied along the $y$-direction, while PMLs are used to truncate the domains above and below the metasurface. Then, the reflectivity ($R$) and transmittivity ($T$) are evaluated, considering only the zeroth-order diffraction channels. In the second configuration, in-plane Bloch wavevector parallel to the metasurface is employed to analyze intrinsic energy radiation into the upper ($w_t$) and lower ($w_b$) half-spaces in the $x$-$y$ plane. The angle $\theta_t$ denotes the angle between the $w_t$ and the $y$-direction, while the $\theta_b$ denotes the angle between the $w_b$ and the $y$-direction.

Figure 2(b) depicts the $R$ and reflected phase difference ($\Delta\phi_E$) spectra of the metasurface as a function of $dP$. The reference plane for $\Delta\phi_E$ is defined at the top surface of the metasurface. The metasurface exhibits an electric mirror response, characterized by $R\geq0.85$ and $\Delta\phi_E\approx0.72\pi$ [40]. Notably, this response remains robust over a wide range of $dP$, indicating strong tolerance to structural variations and suggesting its suitability for disorder-immune electric mirror metasurfaces [38, 39]. Figures 2(c) and 2(d) show the |**E**| distributions within the unit cells at $dP$=0.1λ and 0.3λ, respectively. The suppressed field intensity at the top surface, resulting from destructive interference, further confirms the electric mirror behavior.

Next, we analyze the intrinsic radiation responses of the metasurface under an in-plane Bloch wavevector. Figure 3(a) shows the photonic band diagram of $TE_1$ and $TE_2$ bands as a function of $k_x$, the wavevector along the $x$-direction. The eigen-wavelength of the $TE_2$ band at the Γ point is 1550 nm, indicating that the multilayer cylinder simultaneously excites overlapping MD and EQ resonances. In contrast, the $TE_1$ band exhibits a larger eigen-wavelength. The complex frequency of each band is expressed as $\omega=\omega_0-i\omega_i$, and the dimensionless total quality factor is defined as $Q=\omega_0/2\omega_i$ [50]. Here, $Q$ represents the radiative quality factor only, as dissipative losses are negligible due to the lossless nature of both materials. Figure 3(b) presents the corresponding $Q$ spectra as a function of $k_x$. Notably, the $Q$ of the $TE_2$ band at the Γ point rises sharply to over $10^7$, approaching infinity in theory.

The theoretical intrinsic total radiative loss is given by $\gamma= \omega_0/Q$ [2, 50]. Due to structural symmetry along the $y$-direction, the intrinsic radiative losses toward the top ($\gamma_t$) and bottom ($\gamma_b$) are identical, i.e., $\gamma_t=\gamma_b=\gamma/2$ [2, 51]. Figures 3(c) and 3(d) show the spectra of $\gamma_t$ and $\gamma_b$ for the $TE_2$ band as a function of $k_x$. At the Γ point, both upward and downward radiations decrease sharply by more than two orders of magnitude, whereas significant radiation occurs at off-Γ points, indicating that the metasurface is non-radiative at the Γ point. Figures 3(e) and 3(f) present the radiative energy density spectra of the upward ($w_t$) and downward ($w_b$) for the $TE_2$ band. The energy densities follow the same trend as the radiative losses, confirming the existence of a symmetry-protected BIC at the Γ point. In contrast, Figures 3(g)-3(j) display $\gamma_t$, $\gamma_b$, $w_t$, and $w_b$ spectra for the $TE_1$ band, respectively. Here, the metasurface exhibits significant symmetric radiation, and no non-radiative behavior is observed at the Γ point.

Next, we analyze the polarization properties of the far-field electric field for the $TE_2$ band of the metasurface. The far-field response is obtained using the eigenfrequency solver in COMSOL. The eigenmode is denoted as $\mathbf{E}(k_x, k_z)$, where $k_x$ and $k_z$ are the in-plane wavevectors. Accordingly, the polarization vector of the far-field electric field can be expressed as follows [52]

$$\mathbf{c}(k_x,k_z)=(c_x,c_y,c_z)=\frac{\iint_s e^{ik_x x+k_z z}\mathbf{E}(k_x,k_z)dxdz}{\iint_s dxdz}\ , \tag{8}$$

where $s$ denotes an $x$-$z$ plane slice above the unit cell of the metasurface model. Based on the projected polarization vector ($\boldsymbol{c}_x$, $\boldsymbol{c}_z$), the Stokes parameters for polarizations can be calculated as

$$S_0 = |c_x|^2 + |c_z|^2 ,\ S_1 = |c_x|^2 - |c_z|^2 ,\ S_2 = 2\,\mathrm{Re}(c_x c_z^*) ,\ S_3 = -2\,\mathrm{Im}(c_x c_z^*) . \tag{9}$$

The orientation angle of the polarization state is [53]

$$\varphi(k_x, k_z) = \frac{1}{2}\arg\left[S_1(k_x, k_z) + iS_2(k_x, k_z)\right] . \tag{10}$$

We focus on the region near the Γ point. Figure 4(a) shows the far-field polarization vector map of the $TE_2$ band. Nearly linear polarization states are observed at off-Γ points ($k_x \neq 0$), whereas a pronounced elliptical polarization state emerges at the Γ points ($k_x$=0). A polarization singularity is identified at $k_x$=0 and $k_z$=0, as marked by the green circle in Figure 4(a).

According to the far-field polarization state distributions, the topological charge is calculated by [53, 54]

$$q = \frac{1}{2\pi}\oint_L d\mathbf{k}_{\parallel} \cdot \nabla_{\mathbf{k}_{\parallel}} \varphi(\mathbf{k}_{\parallel}) , \tag{11}$$

$L$ denotes a closed loop in $k$-space surrounding the Γ point, and $\mathbf{k}_{\parallel}$ = ($k_x$, $k_z$) represents the in-plane wavevector. The corresponding topological charge at the singularity in Figure 4(a) is $q$ =1. Figure 4(b) presents the lg($Q$) distribution in the in-plane $k$-space near the Γ point. The off-Γ regions exhibit relatively low $Q$ values, whereas $Q$ increases by more than five orders of magnitude as $k_x \rightarrow 0$, reaching a maximum at the polarization singularity. Figures 4(c) and 4(d) show the electric and magnetic field magnitude distributions within a unit cell at the Γ point. The electromagnetic energy is partially confined in the near-field region around the metasurface and predominantly localized within the cylinders. The field profiles further indicate the excitation of a pronounced EQ resonance, consistent with the scattering analysis above. Together, the coexistence of a far-field polarization singularity, ultrahigh $Q$, and suppressed radiation at the Γ point of the $TE_2$ band confirms the realization of a symmetry-protected BIC in the substrate-free metasurface.

### 2.3 Out-of-plane symmetry-breaking metasurface

For practical optical devices, the influence of the substrate on the device response must be considered. In this work, we design a multilayer substrate with an ABAB stacking sequence to introduce out-of-plane symmetry breaking. Figure 5(a) presents a schematic of the structure, consisting of periodic layers along the $y$-direction and extending infinitely in the transverse directions. Here, layer A (green) corresponds to Si, and layer B (blue) corresponds to $SiO_2$, with thicknesses $t_A$ and $t_B$, respectively. For the unit cell of the $SiO_2$-Si-$SiO_2$ system, the Si layer functions as a high-reflectivity spacer tailored to the BIC operating wavelength of 1550 nm. Under the condition of constructive interference, the minimum thickness of the Si layer satisfies $t_A=\lambda/4n_{Si}$ [55], where the refractive index of Si is approximately 3.48 in the near-infrared regime [56].

For the first type of periodic multilayer films, the thickness ratio satisfies $t_B/t_A = 4.0825$. Figure 5(b) presents the photonic band diagram of the corresponding infinite periodic structure. A photonic bandgap is observed between the lower band 1 and the upper band 2, with a gap-midgap ratio of $\Delta\omega/\omega_m$=49.76%. The green marker denotes the frequency corresponding to a wavelength of 1550 nm, with a normalized wavevector $k_y$=0.36, matching the free-space wavevector at this wavelength. Accordingly, the BIC operating wavelength lies precisely on band 2 at this point, indicating the presence of a localized guided mode. Figure 5(c) demonstrates the |**E**| distribution within a unit cell at the green point. The electromagnetic energy is predominantly confined within the $SiO_2$ layer, whereas the Si layer acts as a high-reflectivity spacer for the 1550 nm and therefore carries negligible field energy.

We next design a practical metasurface comprising periodic cylinders integrated with a multilayer substrate. Figure 6(a) presents a schematic of the unit cell, where the substrate consists of two Si layers and two $SiO_2$ layers, corresponding to a period number of $n_L$=2. Notably, this configuration satisfies the conditions of the first type of periodic multilayer films. Figure 6(b) shows the photonic band diagram of the metasurface. Compared with Figure 3(a), the band structure is significantly modified. Nevertheless, the eigenwavelength of the $TE_2$ band at the Γ point exhibits only a slight redshift to approximately 1550.6 nm. Figure 6(c) presents the corresponding $Q$ spectra of the two bands. The $TE_2$ band has a slightly increased ultrahigh $Q$ at the Γ point, while $Q$ decreases sharply at the off-Γ points, consistent with the behavior observed in Figure 3(b). In contrast, the $TE_1$ band exhibits two $Q$ peaks at normalized wavevectors of ±0.065. Compared with the substrate-free case, the peak values of $Q$ increase by approximately a factor of 6.27.

Figure 6(d) shows the radiation spectra of the $TE_1$ band as a function of $k_x$. The red and blue curves correspond to the $w_t$ and $w_b$, respectively. Notably, two pronounced dips are observed in the $w_t$ spectrum at normalized wavevectors of $k_x \approx \pm 0.071$, as highlighted by the black circles. To quantify the asymmetry between upward and downward radiation, we define a parameter expressed as follows

$$\rho = \frac{w_b}{w_t} \quad . \tag{12}$$

At the black circles, $\rho$ reaches approximately 54.80, indicating strongly unidirectional radiation, with the downward emission significantly exceeding the upward component. These points are located near the peaks of the $TE_1$ band $Q$ spectrum in Figure 6(c). In contrast, the green circles on the blue curve at points of $k_x \approx \pm 0.016$ correspond to minima in downward radiation. Here, $\rho \approx 0.0114$, indicating pronounced unidirectional upward radiation. Figure 6(e) shows the radiation spectra of the $TE_2$ band. Compared with Figure 3(f), the downward radiation decreases over the entire $k$-space range, whereas the upward radiation remains nearly unchanged relative to Figure 3(e). Notably, both upward and downward radiation simultaneously approach zero at the Γ point, resulting in a far-field non-radiative state.

Figure 6(f) shows the spatial distribution of the $x$-component of the electric field ($\mathbf{E}_x$) corresponding to the black circles in Figure 6(d). The field profile clearly demonstrates strongly unidirectional radiation, with the upward emission effectively suppressed and the downward emission remaining dominant. According to momentum conservation, the corresponding wavevector satisfies

$$k_x = |\mathbf{k}| \sin\theta_j \;, (\theta_j = \theta_t,\; \theta_b) \;. \tag{13}$$

The radiation angle $\theta_b$, measured from the surface normal (along the $y$-axis), is approximately 5.13∘. Figure 6(b) further indicates that the corresponding operating wavelength for this unidirectional radiation is 1595.55 nm. Figure 6(g) demonstrates the $\mathbf{E}_x$ field distribution at the green circle. In this case, the radiation is predominantly upward, with the downward component effectively suppressed. The corresponding radiation angle $\theta_t$ is about 1.15∘, with an operating wavelength of 1589.9 nm. Figure 6(h) presents the $\mathbf{E}_x$ distribution at the Γ point of the $TE_2$ band. Here, the metasurface exhibits a far-field non-radiative state, with both upward and downward radiation strongly suppressed. In Figures 6(f)-6(h), the cylinders support pronounced Mie resonances, while the multilayer substrate sustains distinct guided modes. Their interplay gives rise to both unidirectional radiation and BIC responses.

Figure 6(i) presents the far-field polarization distribution in $k$-space around the Γ point of the $TE_2$ band. Elliptically polarized states are observed in the vicinity of $k_x$=0. At the Γ point, a polarization singularity is identified, as marked by the green point. Figure 6(j) shows the lg($Q$) distribution in $k$-space for the $TE_2$ band. The $Q$ reaches its maximum at the Γ point and decreases sharply away from it, confirming the persistence of a symmetry-protected BIC response at 1550 nm, consistent with the substrate-free case. This preserved BIC response can be attributed to the guided mode at 1550 nm, which confines most of the electromagnetic energy within the $SiO_2$ layer. Meanwhile, the high-reflectivity Si layer acts as a spacer, effectively suppressing coupling between the substrate and the cylinders. By contrast, at other wavelengths where these conditions are not satisfied, coupling between the guided mode and the Mie resonances becomes significant, giving rise to pronounced unidirectional radiation.

**2.4 Tunable BIC and directional radiation**

In this section, we investigate the evolution of tunable BIC and directional radiation as the period number ($n_L$) varies. Figure 7(a) presents the eigen-wavelength (λ, blue circle) and $Q$ (red square) of the BIC as a function of $n_L$. λ exhibits a slight blueshift from 1551.17 nm to 1549.44 nm as $n_L$ increases from 1 to 8. A power-law fit yields the relation $\lambda = 10.2173\times n_L^{-0.0935}+1541$ nm. To quantify the wavelength stability, we define the average deviation ratio as

$$\alpha = \frac{1}{N}\sum_{i=1}^{N}\left|\frac{\lambda_i-\lambda_0}{\lambda_0}\right|, \tag{14}$$

where $\lambda_i$ is the BIC wavelength at $n_L=i$ ($i$=1,2,…,8), $\lambda_0$ =1550 nm corresponds to the substrate-free case, and $N$ is the maximum of $n_L$. Here, $\alpha$ is about $3.09\times10^{-4}$, indicating excellent wavelength stability, with a total variation of only $\Delta\lambda$ =1.73 nm over the entire range. Meanwhile, the $Q$ increases monotonically with $n_L$, obeying the trend of $1.802\times10^{6}\times n_L^{(1/3)}+9.136\times10^{6}$. The corresponding average deviation ratio is defined as

$$\beta = \frac{1}{N}\sum_{i=1}^{N}\left|\frac{Q_i-Q_0}{Q_0}\right|, \tag{15}$$

where $Q_i$ is the quality factor at $n_L=i$, and $Q_0$ corresponds to the substrate-free case. The $\beta$ is about 0.1307, confirming the enhancement of the $Q$ with increasing $n_L$.

Figure 7(b) presents the λ and $\theta_b$ spectra of the directional downward radiation as a function of $n_L$. As $n_L$ increases, λ generally exhibits a blueshift with a total variation of 29.28 nm. Meanwhile, $\theta_b$ decreases following the scaling $10.6526^\circ\times n_L^{-1.0882}+0.0161^\circ$ degree, indicating that the radiation direction progressively approaches the surface normal. Figure 7(d) depicts the corresponding $w_b$ and $\rho$. Both $w_b$ and $\rho$ decrease with increasing $n_L$, with a maximum $\rho\approx 54.61$ observed at $n_L$=2.

Figure 7(c) shows the λ and $\theta_t$ of the directional upward radiation as functions of $n_L$. λ exhibits a blueshift described by $85.9381\times n_L^{-1.1794}+1551.6$ nm. In contrast, $\theta_t$ first increases and then decreases, reaching a maximum of 1.21° at $n_L$=3, before gradually approaching the surface normal at larger $n_L$. Figure 7(e) presents the $w_t$ and the inverse asymmetry ratio $\rho^{-1}=w_t/w_b$. The $w_t$ decreases monotonically with increasing $n_L$, while $\rho^{-1}$ exhibits a sharp variation for $n_L\leq4$, reaching a maximum of 87.966 at $n_L$=2, followed by a gradual decline. Overall, the outward directional radiation energy gradually diminishes as $n_L$ increases. This behavior arises because larger $n_L$ introduces more $SiO_2$ layers that confine electromagnetic energy through guided modes, thereby reducing radiation leakage into free space.

For comparison, we consider two additional substrate configurations. The second type shares the same material parameters as the first but adopts a reversed stacking sequence (BABA). This configuration also supports a guided mode at 1550 nm in the corresponding infinite periodic structure. Figure 8(a) shows a schematic of the metasurface incorporating the second-type substrate with $n_L$=2, where the blue and green layers denote $SiO_2$ and Si, respectively. Figure 8(b) illustrates the eigen-wavelength and $Q$ of the BIC at the Γ point as a function of $n_L$. As $n_L$ increases, λ exhibits a pronounced blueshift described by $18.8543\times n_L^{-0.931}+1515.4$ nm. The $\alpha$ = 0.0179 and the total variation Δλ = 15.6 nm (for $n_L$ = 1-8) are both significantly larger than those of the first-type substrate (Figure 7(a)), indicating a higher sensitivity of the eigen-wavelength to $n_L$. Meanwhile, the $Q$ increases according to the trend $4.1225\times10^6\times n_L^{0.2723}+4.394\times10^6$, with $\beta$ = 0.0812, which is smaller than that of the first-type substrate. Although $Q$ increases with $n_L$ in both configurations, the growth rate is reduced in the second type. This behavior arises from enhanced coupling induced by the direct contact between cylinders and the $SiO_2$ layer. As a result, the system exhibits a larger wavelength shift and a slower increase in $Q$.

Finally, we consider the third-type substrate derived from the multilayer films depicted in Figure 8(c). This structure is similar to the first type but with a modified thickness ratio $t_B/t_A$ = 2.40. Figure 8(d) presents the band structure of the corresponding periodic films, exhibiting a photonic bandgap with $\Delta\omega/\omega_m$=54.03%. The green point at the center of the gap corresponds to a wavelength of 1550 nm, with a normalized wavevector $ky$ = 0.24425, matching the intrinsic free-space wavevector along the $y$-direction. Under these conditions, no guided mode is supported at 1550 nm. Figure 8(e) presents the eigen-wavelength and $Q$ of the metasurface incorporating the third-type substrate as a function of $n_L$. As $n_L$ increases, λ exhibits a pronounced blueshift following the scaling $142.5662\times n_L^{-0.649}+1368.2$ nm. The total variation reaches Δλ = 101.16 nm (for $n_L$ = 1-8), with $\alpha$ = 0.0743, both significantly larger than those observed for the first and second substrate types. This indicates a strong sensitivity of the BIC wavelength to variations in $n_L$, accompanied by a significant deviation from the initial 1550 nm. The $Q$ decreases rapidly with increasing $n_L$, following $2.06629\times10^6\times n_L^{-2.324}+1.3711\times10^5$. Here, the $\beta$ = 0.9536 is much larger than the first- and second-type metasurfaces. This behavior originates from the wavelength mismatch between the guided mode and the initial BIC resonance. As a result, electromagnetic energy is not confined within the $SiO_2$ layers at the initial BIC wavelength. Instead, the substrate-induced symmetry breaking enhances coupling between the cylinders and the substrate, enabling substantial energy leakage into the substrate and surrounding free space. Consequently, the quality factor degrades rapidly, and the eigen-wavelength and $Q$ deviate significantly from their original BIC values.

## Conclusion

In summary, we propose a new strategy for achieving tunable high-$Q$ BICs and directional radiation through out-of-plane symmetry breaking in dielectric metasurfaces. We design a $MoSe_2$-$SiO_2$ multilayer cylindrical meta-atom that supports overlapping MD and EQ resonances at 1550 nm. A substrate-free metasurface composed of a periodic array of such cylinders with spatial symmetry exhibits an electric mirror response under normal incidence, which remains robust against variations in the lattice period. Under an in-plane Bloch wavevector, the metasurface supports symmetry-protected BIC at 1550 nm, characterized by ultrahigh $Q$, suppressed radiation, and far-field polarization singularity. To break out-of-plane symmetry, we further introduce three types of Si-$SiO_2$ multilayer substrates. For the first type, the substrate supports a guided mode at 1550 nm, with electromagnetic energy primarily confined within the $SiO_2$ layers, while the high-reflectivity Si layer acts as a spacer that suppresses coupling between the meta-atoms and the substrate. As a result, the metasurface preserves the BIC response near 1550 nm, while the $Q$ factor increases significantly with period number. For the second type, the stacking order of the Si and $SiO_2$ layers is reversed. Although a guided mode at 1550 nm is still supported and energy remains localized in the $SiO_2$ layers, direct contact between the meta-atoms and the $SiO_2$ layer leads to enhanced coupling. Consequently, the metasurface exhibits a pronounced blueshift of the eigen-wavelength and a reduced rate of $Q$-factor enhancement. For the third type, the substrate induces a wavelength mismatch between the guided mode and the BIC. In this case, coupling leads to energy leakage into the substrate and the surrounding space. As a result, the BIC eigen-wavelength undergoes a substantial blueshift with increasing period number, accompanied by a sharp decrease in the $Q$ factor.

Moreover, recent studies have explored the control of BIC responses in dielectric metasurfaces using engineered substrates. Representative examples include perfect-reflection multilayer substrates that support toroidal-dipole Fabry-Pérot BICs for chiral detection [6], high-reflectivity $SiO_2$-$Ta_2O_5$ multilayer substrates that enable asymmetrically coupled BICs for high-$Q$ near-perfect absorption [26], and multilayer substrates that excite degenerate guided modes to induce symmetry-protected BICs [53]. Our work provides a clear and systematic elucidation of the underlying mechanism by which multilayer substrates tune BICs and enable directional radiation in dielectric metasurfaces through out-of-plane symmetry breaking. Importantly, the proposed approach is general and can be extended to a wide range of BIC-based metasurface platforms. Looking forward, dynamically tunable BICs and directional radiation can be further explored in all-dielectric metasurfaces integrated with active multilayer substrates, such as those incorporating phase-change materials [23], liquid crystal coatings [57], or two-dimensional (2D) materials [29]. Overall, our findings highlight the significant potential of multilayer substrates for realizing tunable BICs and directional radiation, and establish a versatile design framework for functional BIC metasurfaces, paving the way toward practical optical applications in sensing, optical communications, lasing, and perfect absorption.

**Acknowledgments**

This work was supported by the National Natural Science Foundation of China (12304425, 12374355, 62375186); Sichuan Science and Technology Program (2024NSFSC1354); the Innovation Team Project of Guangdong General Universities (grant no. 2024KCXTD048).

**Disclosures**

The authors declare no competing financial interest.

**Data availability**

Data may be obtained from the authors upon reasonable request.

**References**

[1] H. Friedrich, D. Wintgen, Interfering resonances and bound states in the continuum, Phys. Rev. A 32(6), 3231-3242 (1985).

[2] X. Yin, J. Jin, M. Soljačić, et al., Observation of topologically enabled unidirectional guided resonances, Nature 580, 467-471 (2020).

[3] C. W. Hsu, B. Zhen, J. Lee, S.-L. Chua, S. G. Johnson, J. D. Joannopoulos, M. Soljačić, Observation of trapped light within the radiation continuum, Nature 499,188-191 (2013).

[4] M.-S. Hwang, H.-C. Lee, K.-H. Kim, K.-Y. Jeong, S.-H. Kwon, K. Koshelev, Y. Kivshar, H.-G. Park, Ultralow-threshold laser using super-bound states in the continuum, Nat. Commun. 12, 4135 (2021).

[5] S. J. Shen, B.-R. Lee, Y. C. Peng, Y. J. Wang, Y.-W. Huang, Y. Kivshar, M. L. Tseng, Dielectric high-*Q* metasurfaces for surface-enhanced deep-UV absorption and chiral photonics, ACS Photonics, 12 (6), 2955-2964 (2025).

[6] C. Li, T. He, X. Rao, C. Feng, J. Zhu, S. Dong, Z. Wei, H. Jiao, Y. Shi, Z. Wang, and X. Cheng, Toroidal dipole Fabry-Perot bound states in the continuum metasurfaces for ultrasensitive chiral detection, Photon. Res. 13(9), 2497-2509 (2025).

[7] M. Liang, L. C. Andreani, A. M. Berghuis, J. L. Pura, S. Murai, H. Dong, J. A. Sánchez-Gil, and J. G. Rivas, Tailoring directional chiral emission from molecules coupled to extrinsic chiral quasi-bound states in the continuum, Photon. Res. 12(11), 2462-2473 (2024).

[8] Q. He, S. Sun, S. Xiao, L. Zhou, High-efficiency metasurfaces: principles, realizations, and applications, Adv. Optical Mater., 6(19), 1800415 (2018).

[9] N. Yu, P. Genevet, M. A. Kats, et al., Light propagation with phase discontinuities: generalized laws of reflection and refraction, Science 334(6054), 333-337 (2011).

[10] N. Muhammad, Z. Su, Q. Jiang, Y. Wang, L. Huang, Radiationless optical modes in metasurfaces: recent progress and applications, Light: Science & Applications, 13,192 (2024).

[11] W. Zhao, S. Wang, Y. Jing, H. Ge, Q. Wang, Y. Zeng, B. Jia, N. Xu, Giant structurally modulated intrinsic chirality of quasi-bound states in continuum, Opt. Lett. 50(5), 1649-1652 (2025).

[12] T. Liu, J. Qiu, M. Qin, X. Tu, H. Qiu, F. Wu, T. Yu, Q. Liu, S. Xiao, High-efficiency infrared upconversion imaging with nonlinear silicon metasurfaces empowered by quasi-bound states in the continuum, Opto-Electron Adv. 9, 250257(2026).

[13] M. Qin, G. Wei, H. Xu, R. Ma, H. Li, W. Gao, J. Liu, F. Wu, Polarization-insensitive and polarization-controlled dual-band third-harmonic generation in silicon metasurfaces driven by quasi-bound states in the continuum, Appl. Phys. Lett. 124(5), 051703 (2024).

[14] W. Wong, X. Huang, O. L. C. Lem, C. Jagadish, H. H. Tan, Near-unity spontaneous emission factor InP surface-emitting lasers based on quasi-bound states in the continuum, Sci. Adv. 11(36), eadx6527 (2025).

[15] Z. Wang, T. Liang, H. Zhang, Ultrahigh-Q quadrupole resonance quasi-BICs excited with different polarized light in an all-dielectric metasurface for refractive index sensing, Optics Communications, 591, 132141 (2025).

[16] S. Lv, F. Hu, W. Luo, H. Xu, L. An, Design of tunable selective light-absorbing metasurfaces driven by intrinsically chiral quasi-bound states in the continuum, Opt. Express 32(17), 30053-30064 (2024).

[17] A. Aigner, T. Possmayer, T. Weber, A. A. Antonov, L. de S. Menezes, S. A. Maier, A. Tittl, Optical control of resonances in temporally symmetry-broken metasurfaces, Nature, 644, 896-902 (2025).

[18] J. C. Zhang, D. P. Tsai, S. W. Pang, Non-local bound states in the continuum for nanoscale alignment, Nat. Photon. 20, 296-300 (2026).

[19] A. Beisenova, W. Adi, W. Wu, S. K. Biswas, S. Rosas, B. Stamenic, D. D. John, F. Yesilkoy, Wafer-scale all-dielectric quasi-BIC metasurfaces: bridging high-throughput deep-UV lithography with nanophotonic applications, Nano Lett. 26, 2059-2067 (2026).
[20] C. Zhou, R. Jin, H. He, J. Huang, G. Li, L. Huang, Robust ultrahigh-Q resonances in tetramer metasurfaces through centroid symmetry protection and area conservation, Light: Science & Applications, 15, 84(2026).
[21] B. Wang, R. Pan, L. Yang, X. Ji, H. Yang, J. Li, Destructive interference mediated topological transitions in bilayer metasurfaces, Phys. Rev. Lett., 136(4), 043801 (2026).
[22] C. Liu, Z. Du, S. Pang, M. Yang, Y. Li, J. Wang, J. Wang, B. Wang, Evolution of circularly polarized states spawned from off-Γ point BICs, Chin. Opt. Lett. 23(12), 123604 (2025).
[23] J. Zeng, Y. Zhou, X. Fu, J. Yang, Y. Chen, W. Hong, Tunable quasi-bound states in the continuum with intrinsic chirality in a phase-change metasurface, Opt. Express 33(13), 27014-27025 (2025).
[24] S.T. Ha, Y. H. Fu, N. K. Emani, Z. Pan, R. M. Bakker, R. Paniagua-Domínguez, A. I. Kuznetsov. Directional lasing in resonant semiconductor nanoantenna arrays, Nature Nanotech. 13, 1042-1047 (2018).
[25] K. Koshelev, A. Bogdanov, Y. Kivshar, Meta-optics and bound states in the continuum, Science Bulletin, 64(12), 836-842 (2019).
[26] Z. Huang, J. Wang, W. Jia, S. Zhang, C. Zhou, All-dielectric metasurfaces enabled by quasi-BIC for high-Q near-perfect light absorption, Opt. Lett. 50(1), 105-108 (2025).
[27] C. Liao, L. Zhou, B. Wen, X. Ye, H. Zhu, J. Yang, G. Li, Z. Zhou, J. Zhou, J. Cai, Defect-insensitive bound states in the continuum in antisymmetric trapezoid metasurfaces in the visible range Nanophotonics, 14(22), 3681-3689 (2025).
[28] Y. Li, S. Han, Y. Ao, Extrinsic and intrinsic chiral dual bound states in the continuum with opposite circular dichroism supported by an all-dielectric metasurface, Opt. Lett. 50(17), 5422-5425 (2025).
[29] S. Ma, S. Wen, X. Mi, H. Zhao, Terahertz optical modulator and highly sensitive terahertz sensor governed by bound states in the continuum in graphene-dielectric hybrid metamaterial, Optics Communications 536, 129398 (2023).
[30] T.-H. Liu, H.-Y. Wu, W.-H. S. Cheng, Quasi-bound state in the continuum of the intracoupled all dielectric coherent metasurface, Nano Letters, 26(1), 288-295 (2026).
[31] X. Wang, J. Liu, Z. Zeng, H. Yuan, Z. Li, Z. He, W. Gong, Z. Wang, Reconfigurable and recyclable low-threshold quasi-BIC lasers via a tunable polymer coating, arXiv:2603.16422.
[32] L. Kühner, F. J. Wendisch, A. A. Antonov, J. Bürger, L. Hüttenhofer, L. de S. Menezes, S. A.Maier, M. V. Gorkunov, Y. Kivshar, A. Tittl, Unlocking the out-of-plane dimension for photonic bound states in the continuum to achieve maximum optical chirality, Light Sci Appl 12, 250 (2023).
[33] S. Li, T. Sang, C. Yang, J. Lu, Y. Wang, Phase-change metasurfaces for dynamic control of chiral quasi-bound states in the continuum, Opt. Lett. 48(24), 6488-6491 (2023).
[34] B. Chen, W. Zhu, H. Jiang, L. Wang, C. Zhou, H. Zhang, Chiral quasi-bound states in the continuum with near-unity and switchable circular dichroism in a photosensitive silicon metasurface, Opt. Lett. 50(10), 3321-3324 (2025).
[35] H. He, F. Lai, Y. Zhang, X. Zhang, C. Tian, X. Li, Y. Wang, S. Xiao, L. Huang, Spectro-polarimetric detection enabled by multidimensional metasurface with quasibound states in the continuum, Opto-Electron. Adv. 8(10), 250015(2025).
[36] F. Wu, M. Luo, J. Wu, C. Fan, X. Qi, Y. Jian, D. Liu, S. Xiao, G. Chen, H. Jiang, Y. Sun, H. Chen, Dual quasibound states in the continuum in compound grating waveguide structures for large positive

and negative Goos-Hanchen shifts with perfect reflection, Phys. Rev. A, 104(2), 023518 (2021).
[37] V. Develay, O. Bahrova, I. Septembre, D. Bobylev, C. Brimont, L. Doyennette, B. Alloing, H. Souissi, E. Cambril, S. Bouchoule, J. Zúñiga-Pérez, D. Solnyshkov, G. Malpuech, T. Guillet, Soliton formation in a bound state in the continuum GaN waveguide polariton laser, arXiv:2512.23368.
[38] H. Song, B. Hong, N. Wang, G. P. Wang, Kerker-type positional disorder immune metasurfaces, Opt. Express 31(15), 24243-24259 (2023).
[39] H. Song, X. Zhang, Y. Sun, G. P. Wang, Manipulating reflection-type all-dielectric non-local metasurfaces via the parity of a particle number, Opt. Express, 32(23), 40467-40487 (2024).
[40] W. Liu, Generalized magnetic mirrors, Phys. Rev. Lett. 119(12), 123902 (2017).
[41] W. Liu and Y. S. Kivshar, Generalized Kerker effects in nanophotonics and meta-optics, Opt. Express 26(10), 13085-13105 (2018).
[42] L. V. R. Marcos, J. I. Larruquert, J. A. Méndez, et al., Self-consistent optical constants of $SiO_2$ and $Ta_2O_5$ films, Opt. Mater. Express 6(11), 3622–3637 (2016).
[43] B. Munkhbat, P. Wrobel, T. J. Antosiewicz, T. O. Shegai, Optical constants of several multilayer transition metal dichalcogenides measured by spectroscopic ellipsometry in the 300-1700 nm range: high index, anisotropy, and hyperbolicity, ACS Photonics 9(7), 2398-2407 (2022).
[44] W. Liu, Superscattering pattern shaping for radially anisotropic nanowires, Phys. Rev. A 96(2), 023854 (2017).
[45] H. Chen and L. Gao, Anomalous electromagnetic scattering from radially anisotropic nanowires, Phys. Rev. A 86(3), 033825 (2012).
[46] A. C. Valero, H. K. Shamkhi, A. S. Kupriianov, et al., Superscattering emerging from the physics of bound states in the continuum, Nat. Commun. 14, 4689 (2023).
[47] C. F. Bohren and D. R. Huffman, Absorption and scattering of light by small particles, (Wiley, New York, 1983).
[48] W. Liu and A. E. Miroshnichenko, Beam steering with dielectric metalattices, ACS Photonics 5(5), 1733-1741 (2018).
[49] COMSOL Multiphysics® v. 6.2. www.comsol.com. COMSOL AB, Stockholm, Sweden.
[50] J. D. Joannopoulos, S. G. Johnson, J. N. Winn, et al., Photonic Crystals: Molding the Flow of Light - Second Edition (REV-Revised, 2), (Princeton University Press, 2008).
[51] H. Song, X. Zhang, J. Wang, Y. Sun, G. P. Wang, Bound state in the continuum and polarization-insensitive electric mirror in a low-contrast metasurface, Opt. Express 32(15), 26867-26883 (2024).
[52] M. Kang, L. Mao, S. Zhang, M. Xiao, H. Xu, C. T. Chan, Merging bound states in the continuum by harnessing higher-order topological charges, Light Sci. Appl. 11, 228 (2022).
[53] P. Hu, J. Wang, Q. Jiang, J. Wang, L. Shi, D. Han, Z. Q. Zhang, C. T. Chan, J. Zi, Global phase diagram of bound states in the continuum, Optica 9(12), 1353-1361 (2022).
[54] Y. Zeng, G. Hu, K. Liu, Z. Tang, C.-W. Qiu, Dynamics of topological polarization singularity in momentum space, Phys. Rev. Lett., 127(17), 176101 (2021).
[55] M. Born and E. Wolf, Principles of Optics: Electromagnetic Theory of Propagation, Interference, and Diffraction of Light, 7th (expanded) Edition, (Publishing House of Electronics Industry, Beijing, 2006).
[56] D. T. Pierce and W. E. Spicer, Electronic structure of amorphous Si from photoemission and optical studies, Phys. Rev. B 5(8), 3017-3029 (1972).
[57] B. Yu, F. Yang, M. Zeng, X. Meng, Z. Qian, Y. Tai, T. Li, Liquid crystal enables extraordinarily precise tunability for a high-Q ultra-narrowband filter based on a quasi-BIC metasurface, Adv. Funct.

Mater. 35(2), 2413098 (2025).

**Figures**

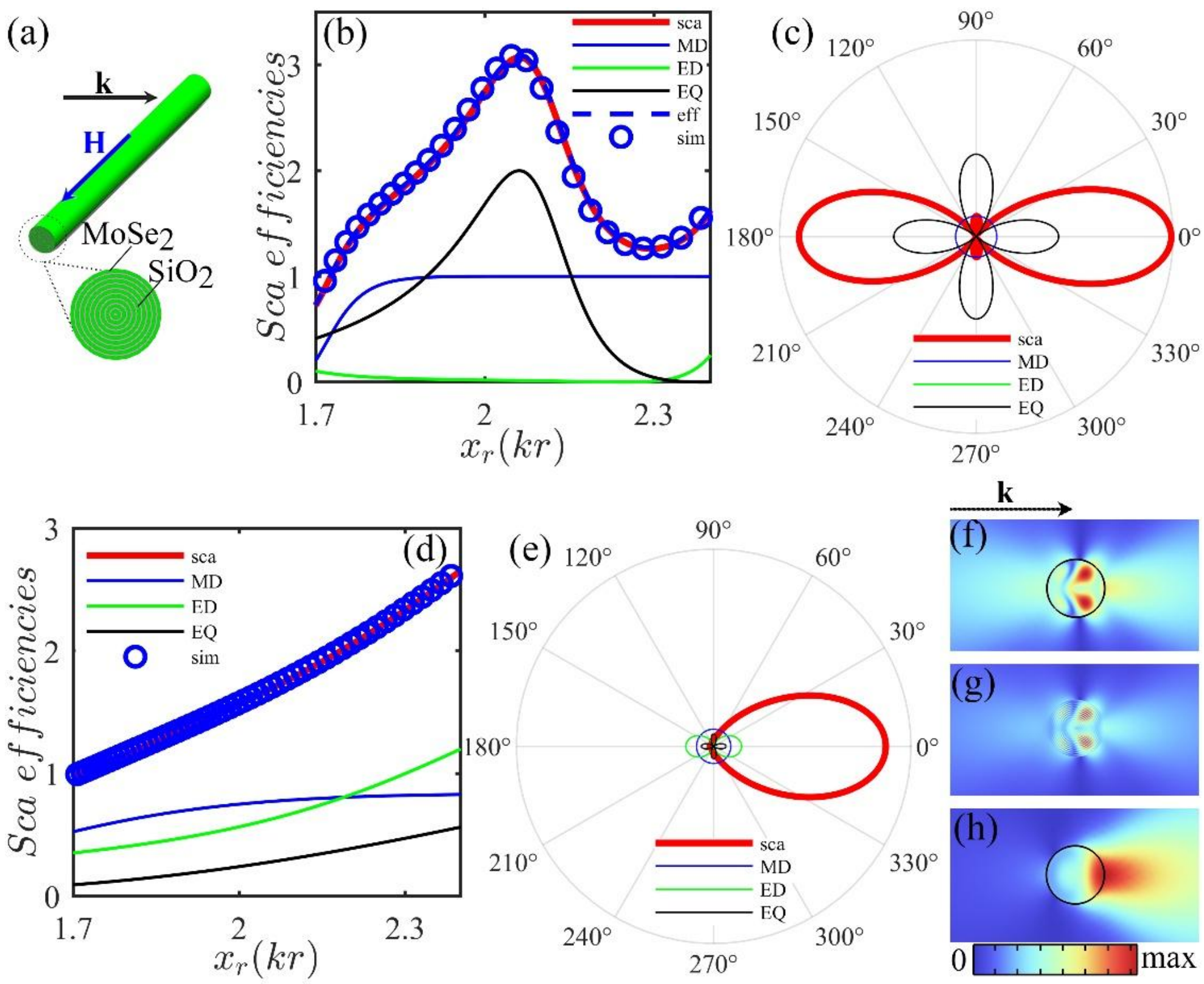

Figure 1. Meta-atom scattering analysis. (a) Schematic of the meta-atom, i.e., a multilayer cylinder. Green denotes the material of $MoSe_2$, gray is $SiO_2$, **K** is the wavevector, and **H** represents the magnetic field. (b) Total scattering efficiency of the cylinder (sca), along with scattering contributions of magnetic dipole (MD), electric dipole (ED), and electric quadrupole (EQ). Blue dashed curve and circle based on the effective parameters (eff) and simulation (sim), respectively. (c) Angular scatterings of the cylinder at $x_r$=2.06. Scattering analysis of a $SiO_2$ cylinder with identical radius, (d) total scattering and multipolar contributions, and (e) angular scattering at $x_r$=2.06. The scattering electrical field of the cylinder at $x_r$=2.06, calculated by (f) effective parameters, (g) $MoSe_2$ and $SiO_2$ multilayer cylinder. (h) The scattering electrical field of the $SiO_2$ cylinder at $x_r$=2.06.

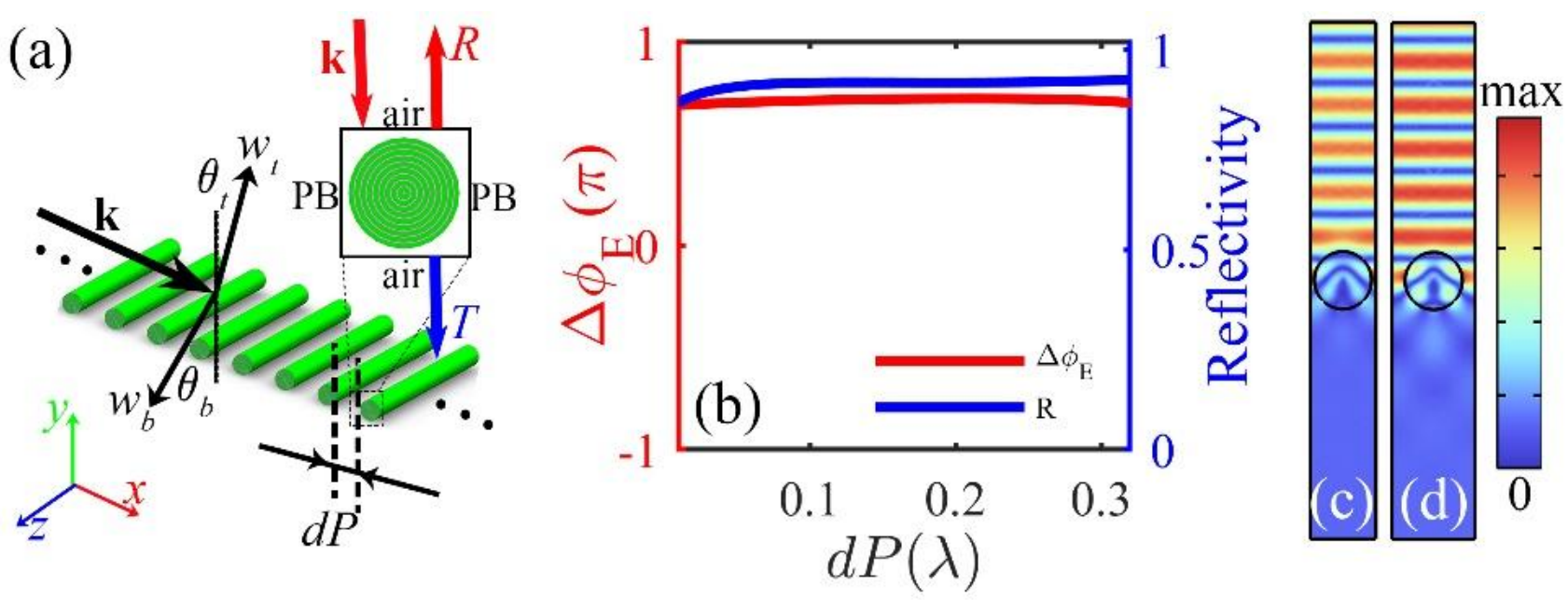

Figure 2. Responses of a metasurface composed of periodic cylinders in free space. (a) Schematic of the metasurface. Reflectivity ($R$) and Transmissivity ($T$) under normal incidence. Upward ($w_t$) and downward ($w_b$) radiation energy when considering the in-plane Bloch wavevector. Upward ($\theta_t$) and downward ($\theta_b$) radiation angles deviating from the $y$-axis in the $x$-$y$ plane. $dP$ denotes the minimal surface distance. (b) $R$ and reflection phase differences spectra as a function of $dP$. (c) and (d) electric field $|\mathbf{E}|$ distributions at $dP$=0.1λ and 0.3λ, respectively.

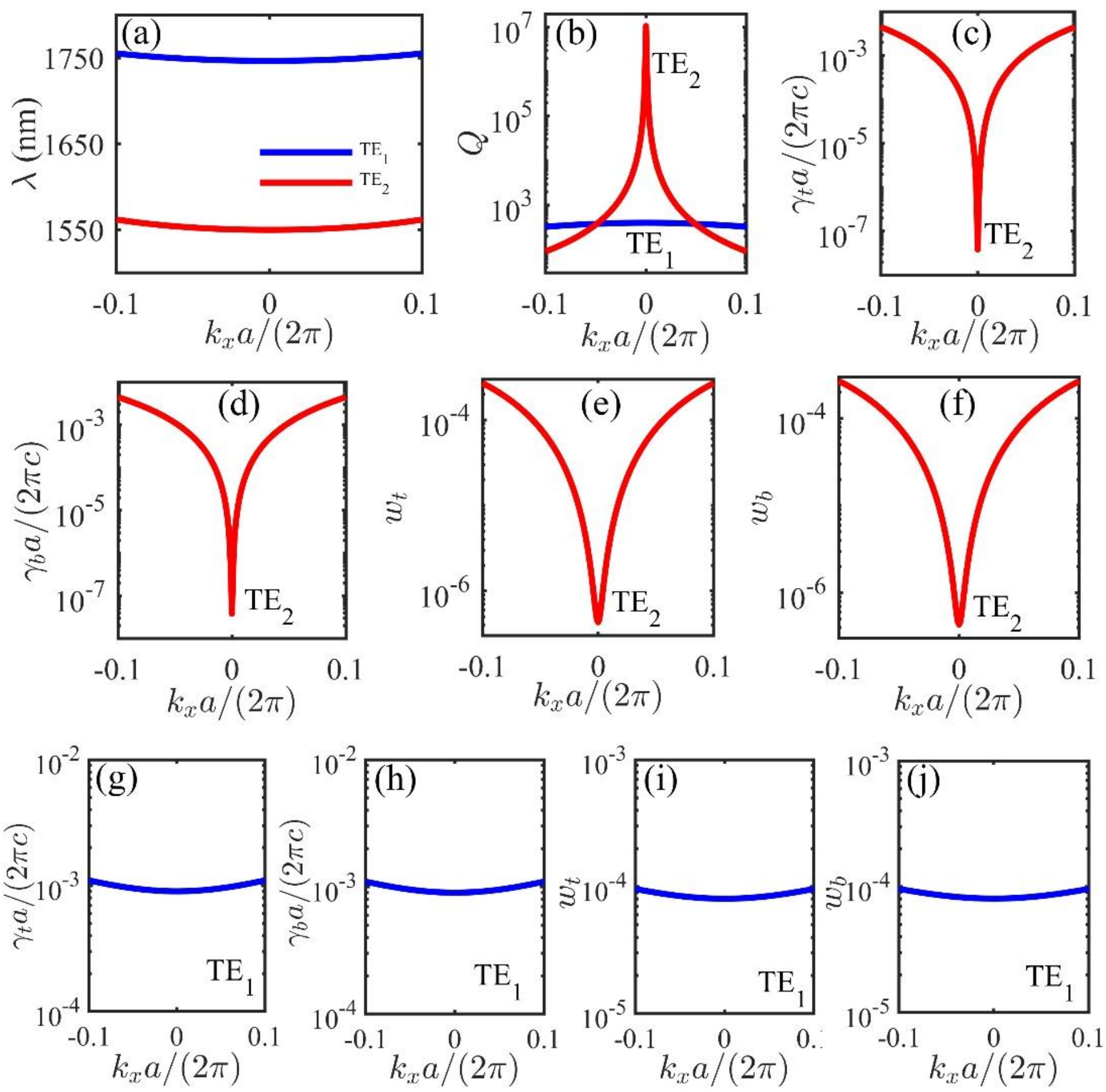

Figure 3. BIC in substrate-free Metasurface. (a) Band diagram of the metasurface with *dP*=0.1615. (b) Quality factor ($Q$) of $TE_1$ and $TE_2$ bands. (c) and (d) Normalized theoretical radiation loss of upward ($\gamma_t$) and downward ($\gamma_b$) of the $TE_2$ band, respectively. (e) and (f) $w_t$ and $w_b$ of the $TE_2$ band, respectively. (g) and (h) Corresponding to normalized $\gamma_t$ and $\gamma_b$ of the $TE_1$ band. (i) and (j) Corresponding to $w_t$ and $w_b$ of the $TE_1$ band.

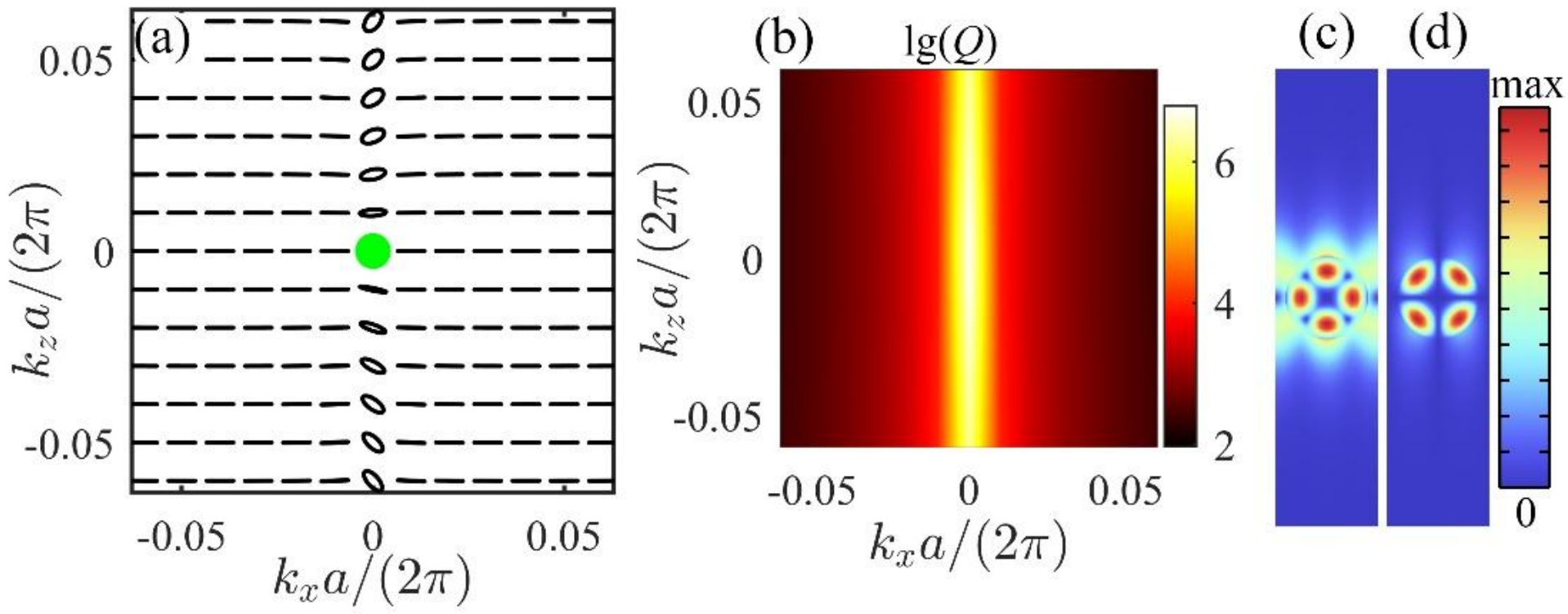

Figure 4. Characterizing BIC properties. (a) Far-field polarization distribution and (b) lg($Q$) in k space. (c) and (d) Electric |**E**| and magnetic field |**H**| distributions at the Γ point, respectively.

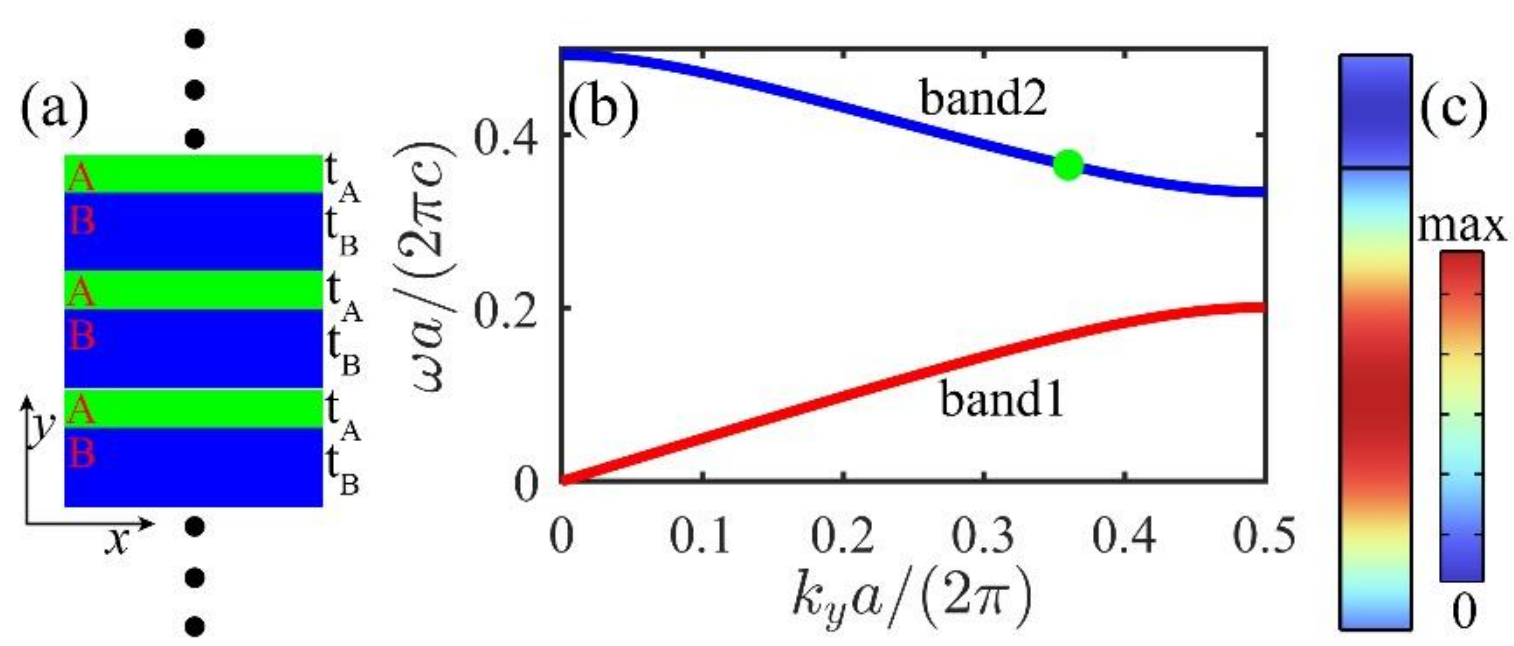

Figure 5. Responses of the first-type one-dimensional (1D) photonic crystal (PC). (a) Schematic of the PC. A layer denotes Si with a thickness of $t_A$, and B is the $SiO_2$ layer with a thickness of $t_B$. (b) Band structure of the PC and green point corresponding to the eigen-wavelength of 1550 nm. (c) |**E**| distribution at the green point.

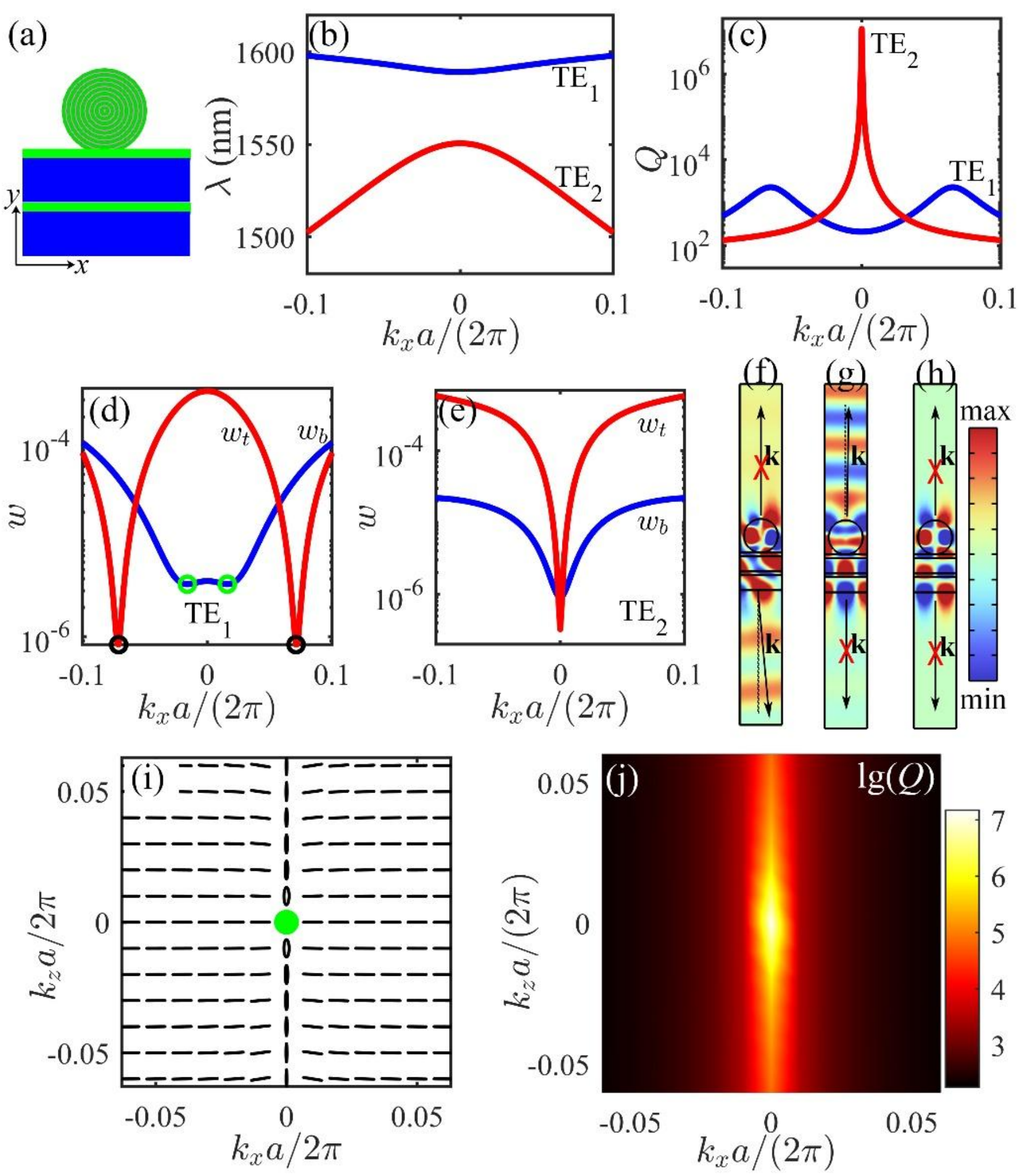

Figure 6. Guided modes tuning BIC and directional radiation. (a) Side view of the unit cell of the metasurface with a four-layer substrate. (b) Band diagram of the metasurface. (c) $Q$ of $TE_1$ and $TE_2$ bands. (d) $w_t$ and $w_b$ spectra of the $TE_1$ band. (e) $w_t$ and $w_b$ spectra of the $TE_2$ band. (f)-(h) Electric field distributions ($\mathbf{E}_x$) at the black circle, green circle, and Γ points, respectively. (i) Far-field polarization distribution of the $TE_2$ band. (j) lg($Q$) spectra of the $TE_2$ band.

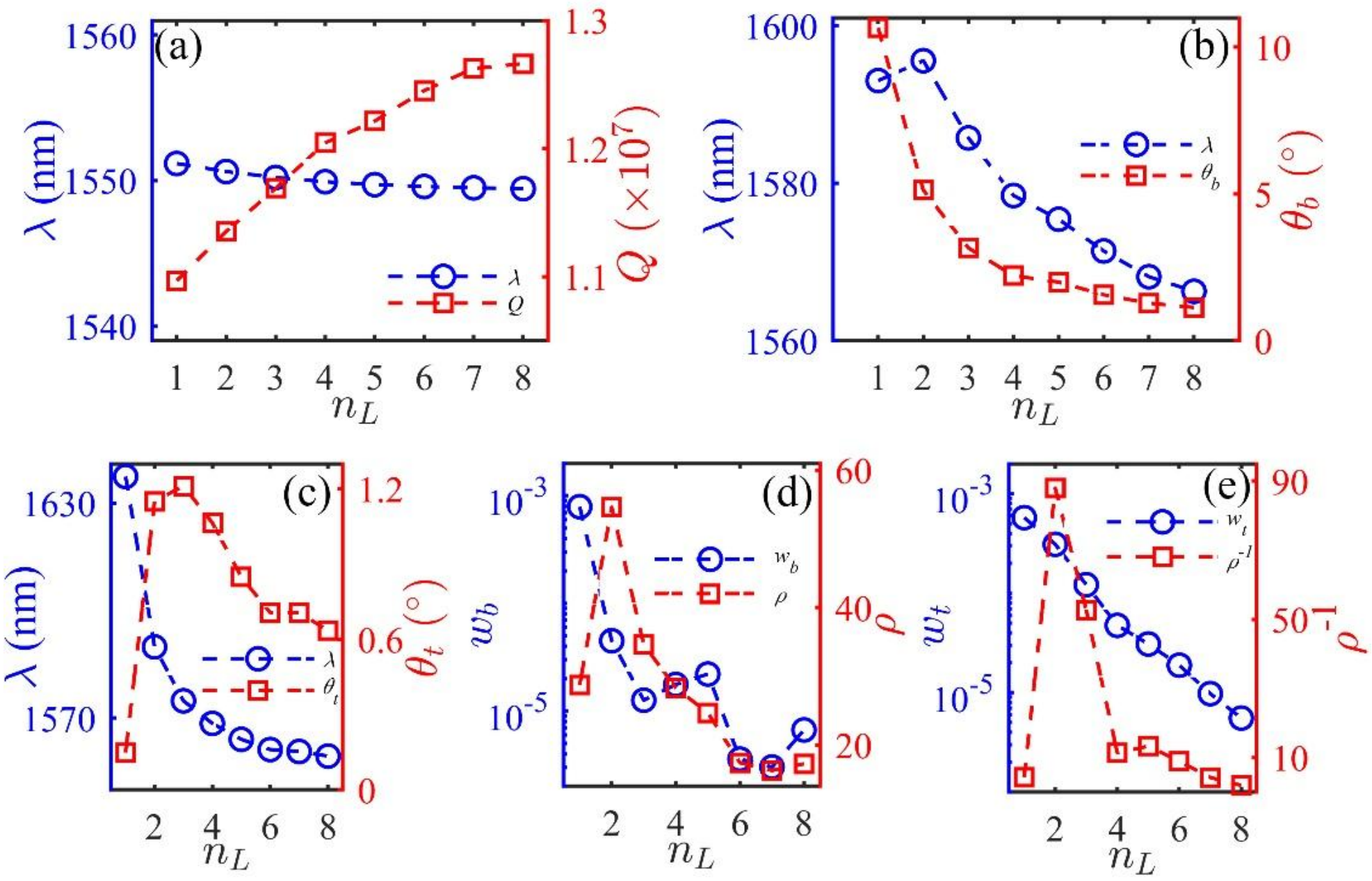

Figure 7. Period number $n_L$-dependent tunable BIC and directional radiation in the metasurface with the first-type substrate. (a) Eigen-wavelength and $Q$ spectra of the BIC. (b) Wavelength and $\theta_b$ spectra of the downward radiation. (c) Wavelength and $\theta_t$ spectra of the upward radiation. (d) The $w_b$ and $\rho$ spectra of the downward radiation. And, $\rho = w_b / w_t$. (e) The $w_t$ and $\rho$ spectra of the upward radiation.

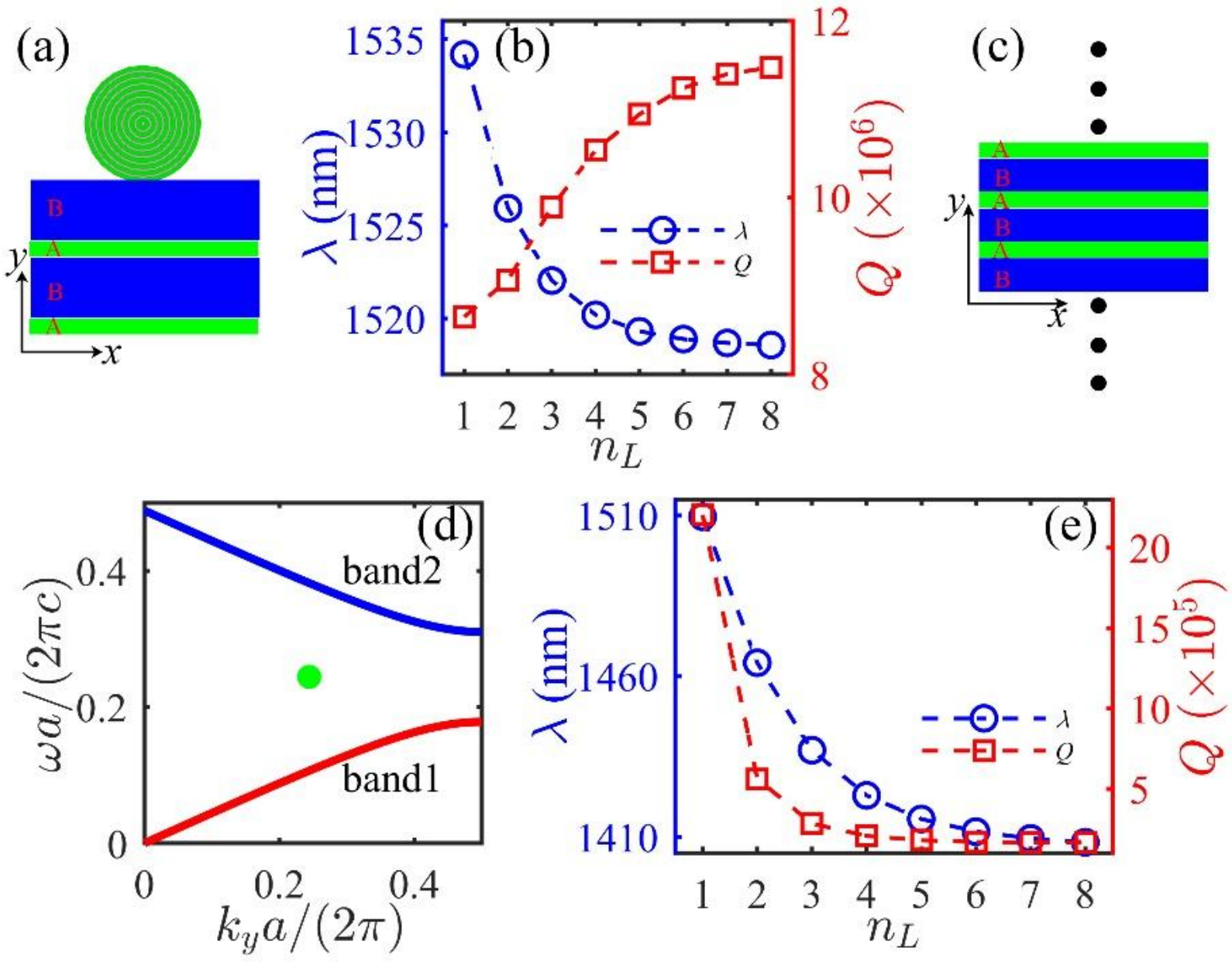

Figure 8. BIC responses comparison for the metasurfaces with the second-type and third-type substrates, respectively. (a) Schematic of the metasurface with the second-type substrate. (b) Eigen-wavelength and *Q* of BIC spectra as a function of $n_L$ for the metasurface in (a). (c) Schematic of another 1D PC with a smaller $t_B$. The third-type substrate is composed of this finite PC. (d) Band structure of the PC. Green point corresponding to 1550 nm. (d) Eigen-wavelength and *Q* of BIC spectra concerning $n_L$ for the metasurface with the third-type substrate.